\documentclass[12pt]{article}

\usepackage{graphicx}
\usepackage{amsmath,amsfonts}
\usepackage{bm}
\newcommand{\uu}[1]{{#1}}
\def\calI{{\mathcal I}}
\newcommand{\jmp}[1]{[\![#1]\!]}
\newcommand{\mvl}[1]{\{\!\!\{#1\}\!\!\}}
\def\hh{{\tt h}}
\def\cc{{\tt c}}
\def\pp{{\tt p}}
\title{Dynamical energy analysis for built-up acoustic systems at high frequencies}

\author{D.J. Chappell\footnote{david.chappell@nottingham.ac.uk}, S. Giani\footnote{stefano.giani@nottingham.ac.uk} and G. Tanner\footnote{gregor.tanner@nottingham.ac.uk}\\
School of Mathematical Sciences, University of Nottingham,\\
University Park, Nottingham NG7 2RD, UK}
\date{}
\begin{document}
\maketitle

\begin{abstract}
Standard methods for describing the intensity distribution of mechanical and acoustic wave fields in the high frequency asymptotic limit are often based on flow transport equations.
Common techniques are statistical energy analysis, employed mostly in the context of vibro-acoustics, and ray tracing, a popular tool in architectural acoustics.
Dynamical energy analysis makes it possible to interpolate between standard statistical energy analysis and full ray tracing, containing both of these methods as limiting cases.
In this work a version of dynamical energy analysis based on a Chebyshev basis expansion of the Perron-Frobenius operator governing the ray dynamics is introduced.
It is shown that the technique can efficiently deal with multi-component systems overcoming typical geometrical limitations present in statistical energy analysis.
Results are compared with state-of-the-art hp-adaptive discontinuous Galerkin finite element simulations.
\end{abstract}


\section{INTRODUCTION}

Predicting the wave energy distribution of the vibro-acoustic response of a complex mechanical system to periodic excitation is a challenging task, especially in the mid-to-high frequency regime. Standard numerical tools such as finite element methods become inefficient, and ray or thermodynamic approaches are often employed to model the wave energy flow through the structure. Popular methods are \textit{Statistical Energy Analysis} (SEA) \cite{RL69,RL95, RC96}, in which the mean energy flow between subsystems is assumed to be proportional to the energy gradient, and the \textit{ray tracing technique}, in which the wave intensity distribution is determined by summing over contributions of a potentially large number of ray paths \cite{AG89, HK00, VC01}.

Ray tracing and SEA both predict mean values of the energy distribution and omit information about wave effects such as interference or diffraction. Both methods are therefore expected to hold in the high frequency (or small wavelength) limit. SEA is in fact a low resolution ray tracing method \cite{SK01,GT09} leading to small numerical models compared to ray tracing. This efficiency saving comes at a price, however: SEA has no spatial resolution of the energy distribution within subsystems and becomes unreliable whenever long range correlations in the ray dynamics are present. The recently developed \textit{Dynamical Energy Analysis} (DEA) \cite{GT09} provides a tool which interpolates between SEA and a full ray tracing analysis and can overcome some of the problems mentioned above at a relatively small computational overhead. DEA thus enhances the range of applicability of standard SEA and gives bounds on the range of applicability of SEA. Related methods have been discussed previously in the context of wave chaos \cite{GT07} and structural dynamics \cite{KH94}. In particular Langley's \textit{Wave Intensity Analysis} (WIA) \cite{RL92, RL94} and Le Bot's thermodynamical high frequency boundary element method \cite{AL98, AL02} include details of the underlying ray dynamics. The approach employed here differs from these methods by considering multiple reflections in terms of linear operators. Representing these operators in terms of basis function expansions then leads to SEA-type equations.

In this work a reformulation of DEA is presented, which is based on a Chebyshev basis function representation; this leads to considerable improvements compared to previous attempts based on an expansion in terms of a Fourier basis \cite{GT09}. Both Chebyshev and Fourier basis expansions of smooth functions share similar exponential convergence properties \cite{JB00}. The main advantages of using a representation in terms of Chebyshev polynomials include that the requirement for periodic boundary conditions can be dropped, allowing for much more freedom in the choice of approximation regions. In addition, a Chebyshev expansion gives rise to more efficient quadrature rules for numerically calculating the arising integrals when constructing the linear operators considered in DEA. Using a Chebyshev basis leads to a natural choice of quadrature, namely Gauss-Chebyshev, which is optimal for polynomial-type integrands and naturally incorporates the orthogonality weighting term in the Chebyshev basis function representation. In order to take full advantage of this feature it is necessary to formulate the problem in terms of the final position and momentum of a given ray, and map back to its initial point. This is in contrast to previous work on DEA by Ref. \cite{GT09}, where the rays were defined by their endpoints. The strengths of the newly reformulated DEA are evident in the applications considered, where due to improved efficiency it has been possible to model multi-component systems with variable wave-numbers for the first time. The method is verified numerically by comparing DEA results with state-of-the-art finite element software for a range of parameter values.

The remainder of the paper is structured as follows. In Section \ref{sec:green}, the ray tracing approximation is discussed and related to the Green function using short wavelength asymptotics. In Section \ref{sec:PF}, the concept of phase-space operators is introduced and their representation in terms of boundary basis functions is discussed. In Section \ref{sec:numericimp} the implementation of DEA is detailed along with its links with SEA. The finite element formulation used for verification of the results is also briefly discussed. In Section \ref{sec:computations}, a variety of coupled two-cavity configurations are discussed and the results compared against finite element computations. Finally larger multi-cavity configurations are considered.

\section{\label{sec:green}WAVE ENERGY FLOW IN TERMS OF THE GREEN FUNCTION}

It is assumed that the system as a whole is characterized by a linear wave equation describing the overall wave dynamics including damping and radiation in a finite domain $\Omega\subset\mathbb{R}^{d}$, $d=2$ or 3. In this work only stationary problems with continuous, monochromatic energy sources are considered. We split the system into $N_{\Omega}$ subsystems and consider the scalar wave equation for acoustic pressure waves in each homogeneous sub-domain $\Omega_{i}$, with local wave velocity $c_{i}$, $i=1,...,N_{\Omega}$ and $\Omega=\bigcup_{i=1}^{N_{\Omega}}\Omega_{i}$. Extensions to more complicated systems with different wave operators in different parts of the system can be treated with the same techniques as long as the underlying wave equations are linear, see the discussion in Ref. \cite{GT09}.

The general problem of determining the response of a system to external forcing with angular frequency $\omega$ at a source point $r_{0}\in\Omega_{0}$ can then be reduced to solving
\begin{equation}\label{wave-eq}
(k_{i}^{2}-\hat{H})G(r,r_{0};\omega)=-\delta(r-r_{0}),\hspace{5mm}i=1,...,N_{\Omega},
\end{equation}
with $\hat{H} = -\Delta$, $G$ represents the Green function, $r\in\Omega_{i}$ is the solution point and $\delta$ is the Dirac delta distribution.
Furthermore, $k_i = \omega/c_i + i \mu_i/2$ is a complex valued wavenumber,
where the imaginary part represents a subsystem dependent damping coefficient
$\mu_{i}$. Throughout this work we take $i=\sqrt{-1}$ unless used as a
subscript, in which case it is an index over the number of subsystems.
The wave energy density induced by the source is then given as
\begin{equation}\label{en-eq1}
\varepsilon(r, r_{0}; \omega)=\frac{|G(r,r_{0};\omega)|^{2}}{\varrho_{i}c_{i}^{2}},
\end{equation}
for $r \in \Omega_i$ where $\varrho_{i}$ is the density of the medium in $\Omega_{i}$.
The linear wave operator $\hat{H}$ can naturally be associated with the underlying ray dynamics via the Eikonal approximation; for more detailed derivation, see Ref. \cite{GT09, OR07, PC09PV}. Using small wavelength asymptotics, the Green function in equation (\ref{wave-eq}) may be written as a sum over \textit{all} classical rays from $r_{0}$ to $r$ for fixed kinetic energy of the hypothetical ray particle. One obtains \cite{MG90, PC09PV}
\begin{equation}\label{Grn-atc}
G(r,r_{0};\omega)=\frac{\pi}{(2\pi i)^{(d+1)/2}}\sum_{j:r_{0}\rightarrow r}A_{j}e^{i(k_{i}R_{j}-i\nu_{j}\pi/2)},
\end{equation}
where $R_{j}$ is the length of the ray trajectory between $r_{0}$ and $r$ including possible reflections on boundaries. The amplitudes $A_{j}$ may be written as a product of three terms as in Ref. \cite{GT09} due to damping, mode conversion and reflection/transmission coefficients, and geometrical factors. The phase index $\nu_{j}$ contains contributions from the reflection/transmission coefficients at interfaces between subsystems and from caustics along the ray path.

Analogous representations to (\ref{Grn-atc}) have been considered in detail in quantum mechanics \cite{MG90} and are also valid for general wave equations in elasticity, see Ref. \cite{GT07} for an overview. In the latter case $G$ becomes matrix valued. Note that the summation in equation (\ref{Grn-atc}) is typically over infinitely many terms, where the number of contributing rays increases (in general) exponentially with the length of the trajectories included. This gives rise to convergence issues, especially in the case of low or no damping \cite{GT07}.

The wave energy density (\ref{en-eq1}) can now be expressed as a double sum over classical trajectories and hence
\begin{equation}\label{en-eq2}
\begin{array}{ll}
{\varepsilon(r, r_{0}; \omega)} & {\displaystyle =C\sum_{j,\: j': r_{0}\rightarrow r}A_{j}A_{j'}e^{ik_{i}[R_{j}-R_{j'}]-i[\nu_{j}-\nu_{j'}]\pi/2}}\vspace{2mm}\\
{} & {=C[\rho(r,r_{0};\omega)+\textrm{off-diagonal\:terms}],}
\end{array}
\end{equation}
with $C=\pi^{2}/(\varrho_{i}c_{i}^{2}(2\pi)^{(d+1)})$. The dominant contributions to the double sum arise from terms in which the phases cancel exactly; one thus splits the calculation into a diagonal part
\begin{equation}\label{rho-def}
\rho(r,r_{0};\omega)=\sum_{j:r_{0}\rightarrow r}|A_{j}|^{2}
\end{equation}
where $j=j'$, and an off-diagonal part. The diagonal contribution gives a smooth background signal and the off-diagonal terms give rise to fluctuations on the scale of the wavelength. The phases related to different trajectories are (largely) uncorrelated and the resulting net contributions to the off-diagonal part are in general small compared to the smooth part, especially when averaging over frequency intervals of a few wavenumbers.

It has been shown in Ref. \cite{GT09} that calculating the smooth diagonal part (\ref{rho-def}) is equivalent to a ray tracing treatment. That is, the smooth part of the energy density can be described in terms of the flow of fictitious non-interacting particles emerging from the source point $r_{0}$ uniformly in all directions and propagating along ray trajectories. This makes it possible to relate wave energy transport with classical flow equations and thus thermodynamical concepts, which are at the heart of an SEA treatment. In DEA the classical flow is expressed in terms of linear \textit{phase space operators} as detailed in the next section.

\section{\label{sec:PF}LINEAR PHASE SPACE OPERATORS AND DEA}

\subsection{Phase space operators and boundary maps}
A brief outline of the derivation of the DEA flow equations is now given, for details see Ref. \cite{GT09}. We adopt a purely kinetic viewpoint based on the interpretation that rays are trajectories of particles following Hamiltonian dynamics as detailed in Section 2 of Ref. \cite{OR07}. Here the time dependence of a density of ray trajectories (or particles) $\tilde{\rho}$ is known to satisfy the Liouville equation
\begin{equation}\label{liouville}
\frac{\partial\tilde{\rho}}{\partial\tau}(X,\tau)+\frac{d X}{d\tau}\cdot\nabla_{X}(\tilde{\rho}(X,\tau))=0,
\end{equation}
where $X=(r,p)$ denotes the phase space coordinate with position $r$ and momentum $p$.
The propagator for the Liouville equation is the linear phase space operator $\mathcal{L}^{\tau}(X,Y)=\delta(X-\varphi^{\tau}(Y))$, known as a Perron-Frobenius operator in dynamical systems theory \cite{PC09C16}, and hence we may write
\begin{equation}\label{fpo1}
\tilde{\rho}(X,\tau)=\int_{\mathbb{P}}\mathcal{L}^{\tau}(X,Y)\tilde{\rho}_{0}(Y)dY.
\end{equation}
Here the phase space flow $\varphi^{\tau}(Y)$ gives the position of the particle after time $\tau$ starting at $Y=(r',p')$ when $\tau=0$. Furthermore, $\tilde{\rho}_{0}$ denotes the initial ray density at time $\tau=0$ and the domain of integration is over the whole of phase space $\mathbb{P}=\Omega\times \mathbb{R}^d$, where the integration over $\mathbb{R}^{d}$ takes care of the momentum coordinates $p$.

Consider a source localized at a point $r_{0}$ emitting waves continuously at a fixed angular frequency $\omega$. Standard ray tracing techniques estimate the wave energy at a receiver point $r$ by determining the density of rays starting at $r_{0}$ and reaching $r$ after some unspecified time. This may be written in the form
\begin{equation}\label{fpo2}
\rho(r,r_{0},\omega)=\small{\int_{0}^{\infty}\int_{\mathbb{R}^d}\int_{\mathbb{P}}} w(Y,\tau)\mathcal{L}^{\tau}(X,Y)\rho_{0}(Y,\omega)dYdp\:d\tau,
\end{equation}
with initial density $\rho_{0}(Y,\omega)=\delta(r'-r_{0})\delta(k_0^{2}-H(Y))$, where $H=|p|^2$ is the Hamilton function for the wave operator $\hat{H}$ and $k_0$ is the wave number at the source point as defined in Eqn.\ (\ref{wave-eq}). It can be shown that equation (\ref{fpo2}) is equivalent to the diagonal approximation (\ref{rho-def}) \cite{GT09}. A weight function $w$ is included to incorporate damping and reflection/transmission coefficients. It is assumed that $w$ is multiplicative, ($w(\varphi^{\tau_{1}}(X),\tau_{2})w(X,\tau_{1})=w(X,\tau_{1}+\tau_{2})$), which holds for standard absorbtion mechanism and reflection processes \cite{PC09PV}.

In order to solve the stationary flow problem (\ref{fpo2}) a boundary mapping technique is employed. For the time being let us consider a problem with a single (sub-)system $\Omega=\Omega_{1}$ with boundary $\Gamma$. The boundary mapping procedure involves first mapping the ray density emanating continuously from the source onto the boundary $\Gamma$. The resulting boundary layer density $\rho_{\Gamma}^{(0)}$ is equivalent to a source density on the boundary producing the same ray field in the interior as the original source field after one reflection. Secondly, densities on the boundary are mapped back onto the boundary by a boundary operator $\mathcal{B}(X^{s},Y^{s};\omega)=w(Y^{s})\delta(X^{s}-\phi^{\omega}(Y^{s}))$, where $X^{s}=(s,p_{s})$ represents the coordinates on the boundary. That is, $s$ parameterizes $\Gamma$ and $p_{s}\in B^{d-1}_{|p|}$ denotes the momentum component tangential to $\Gamma$ at $s$ for fixed $H(X) = |p|^2$, where $B^{d-1}_{|p|}$ is an open ball in $\mathbb{R}^{d-1}$ of radius $|p|$ and centre $s$. Likewise, $Y^{s}=(s',p_{s}')$ and $\phi^{\omega}$ is the invertible boundary map. Note that convexity is assumed to ensure $\phi^{\omega}$ is well defined; non-convex regions can be handled by introducing a cut-off function in the shadow zone as in Ref. \cite{AL02} or by subdividing the regions further.

The stationary density on the boundary induced by the initial boundary distribution $\rho_{\Gamma}^{(0)}(X^{s},\omega)$ can then be obtained using
\begin{equation}\label{Neum-ser}
\rho_{\Gamma}(\omega)=\sum_{n=0}^{\infty}\mathcal{B}^{n}(\omega)\rho_{\Gamma}^{(0)}(\omega)=(I-\mathcal{B}(\omega))^{-1}\rho_{\Gamma}^{(0)}(\omega),
\end{equation}
where $\mathcal{B}^{n}$ contains trajectories undergoing $n$ reflections at the boundary. The resulting density distribution on the boundary $\rho_{\Gamma}(X^{s},\omega)$ can then be mapped back into the interior region. One obtains the density (\ref{fpo2}) after projecting down onto coordinate space.

\subsection{Chebyshev basis representation}
The long term dynamics are thus contained in the operator $(I-\mathcal{B})^{-1}$ and standard properties of Perron-Frobenius operators ensure that the sum over $n$ in equation (\ref{Neum-ser}) converges for non-vanishing dissipation. In order to evaluate $(I-\mathcal{B})^{-1}$ it is convenient to express the operator $\mathcal{B}$ in a suitable set of basis functions defined on the boundary. In Ref. \cite{GT09} a Fourier basis has been applied, which is a natural choice of a complete basis for problems with periodic boundary conditions. However, a number of difficulties arise with this choice such as slow convergence of
quadrature rules for the associated integrals and the treatment of corners on the boundary.

Here we employ a Chebyshev polynomial basis representation with Gauss-Chebyshev quadrature, in which case the integration is optimal for polynomial-type integrands. Problems due to singular behavior at corners are avoided due to integrating over phase space, rather than over pairs of boundary coordinates. Gauss-Chebyshev quadrature incorporates the orthogonality weight functions for the Chebyshev basis automatically in the quadrature rule. In the case $d=2$, we have $s\in[0,L)$ and Chebyshev basis may be expressed in the form
\begin{equation}\label{Cheb-basisg}
\tilde{T}_{n}(X^{s})=\sqrt{\frac{2}{|p|L}}T_{n_{1}}\left(\frac{2s}{L}-1\right)T_{n_{2}}\left(\frac{p_{s}}{|p|}\right),
\end{equation}
with $n=(n_{1},n_{2})$ non-negative integers and $T_{n_{1}}$ the Chebyshev polynomial of order $n_{1}$. The Chebyshev basis approximation $B^{mn}$ of $\mathcal{B}$ may be written:
\begin{equation}\label{cheb1}
\begin{array}{l}
{B^{mn}=}\vspace{2mm}\\
{\displaystyle {\small\int_{\partial\mathbb{P}}\int_{\partial\mathbb{P}}} W_{m}(X^{s})\tilde{T}_{m}(X^{s})\mathcal{B}(X^{s},Y^{s};\omega)\tilde{T}_{n}(Y^{s})dY^{s}dX^{s}}\vspace{2mm}\\
{\displaystyle=\int_{\partial\mathbb{P}} W_{m}(\phi^{\omega}(Y^{s}))\tilde{T}_{m}(\phi^{\omega}(Y^{s}))w(Y^{s})\tilde{T}_{n}(Y^{s})dY^{s},}
\end{array}
\end{equation}
where $\partial\mathbb{P}=[0,L)\times(-|p|,|p|)$ is the phase space on the boundary at fixed ``energy"
$H(X) = |p|^2$ and
\begin{equation}\label{Cheb-weightg}
W_{m}(X^{s})=\frac{4\gamma_{m_{1}}\gamma_{m_{2}}}{\pi^{2}}\frac{1}{\sqrt{1-((2s/L)-1)^{2}}}\frac{1}{\sqrt{1-(p_{s}/|p|)^{2}}},
\end{equation}
is the weight function for the inner product in which the Chebyshev basis is orthonormal. Here $\gamma_{0}=1/2$ and $\gamma_{n_{1}}=1$ for $n_{1}=1,2,...$. It is convenient for the Gauss-Chebyshev quadrature if the argument in the weight function $W_{m}$ is the same as the integration variable and so a change of variables $Y^{s}=\psi^{\omega}(X^{s})$ with $\psi^{\omega}=(\phi^{\omega})^{-1}$ is carried out to give
\begin{equation}\label{cheb2}
\begin{array}{l}
{ B^{mn}=}\\
{\displaystyle\int_{\partial\mathbb{P}} W_{m}(X^{s})\tilde{T}_{m}(X^{s})w(\psi^{\omega}(X^{s}))\tilde{T}_{n}(\psi^{\omega}(X^{s}))|J(X^{s})|dX^{s}.}
\end{array}
\end{equation}
Here the Jacobian term is
\begin{equation}\label{chebJ}
{|J(X^{s})|} {=|\partial\psi^{\omega}(X^{s})|} =\left|
\begin{array}{ll}{\frac{\partial s'}{\partial s}} & {\frac{\partial s'}{\partial p_{s}}}\\
{\frac{\partial p_{s}'}{\partial s}} & {\frac{\partial p_{s}'}{\partial p_{s}}}
\end{array}\right|,
\end{equation}
which is equal to one for Hamiltonian flows and takes account of changes in the wavenumber between subsystems. In this representation the integration is with respect to position and momentum at the end of the ray being considered, and we are mapping back to the initial point using $\psi^{\omega}$.

\subsection{Subsystems}
Recall the splitting into subsystems $\Omega_{i}$, $i=1,..,N_{\Omega}$ introduced earlier. The dynamics in each subsystem are considered separately so that both variability in the wave velocity $c_{i}$ and non-convex domains may be handled simply. Coupling between sub-elements can then be treated as losses in one subsystem and source terms in another. Typical subsystem interfaces are surfaces of reflection/ transmission due to sudden changes in material parameters or local boundary conditions. We describe the full dynamics in terms of subsystem boundary operators $\mathcal{B}_{ij}$; flow between $\Omega_{j}$ and $\Omega_{i}$ is only possible if $\Omega_{i}\cap\Omega_{j}\neq\emptyset$ and one obtains
\begin{equation}\label{fpo-ij}
\mathcal{B}_{ij}(X_{i}^{s},X_{j}^{s})=w_{ij}(X_{j}^{s})\delta(X_{i}^{s}-\phi^{\omega}_{ij}(X_{j}^{s})),
\end{equation}
where $\phi_{ij}^{\omega}$ is the boundary map in subsystem $j$ mapped onto the boundary of the adjacent subsystem $i$ and $X_{i}^{s}$ are the boundary coordinates of $\Omega_{i}$. The weight $w_{ij}$ contains, among other factors, reflection and transmission coefficients characterizing the coupling at the interface between $\Omega_{j}$ and $\Omega_{i}$.

A basis function representation $B^{mn}_{ij}$ of the full operator $\mathcal{B}$ as suggested in equation (\ref{cheb1}) is now written in terms of subsystem boundary basis functions $\tilde{T}^{i}_{n}$ to give
\begin{equation}\label{cheb-ij}
B^{mn}_{ij}=\small{\int_{\partial\mathbb{P}_{i}}\int_{\partial\mathbb{P}_{j}}} W_{m}(X_{i}^{s})\tilde{T}^{i}_{m}(X_{i}^{s})\mathcal{B}(X_{i}^{s},X_{j}^{s};\omega)\tilde{T}^{j}_{n}(X_{j}^{s})dX_{j}^{s}dX_{i}^{s}.
\end{equation}
Here $\partial\mathbb{P}_{i}$ is simply the boundary coordinate phase space for $\Omega_{i}$. The equilibrium distribution on the interfaces of the subsystems is then obtained by solving the system of equations (\ref{Neum-ser})
\begin{equation}\label{lin-syst}
(I-B)\rho_{\Gamma}=\rho_{\Gamma}^{(0)}.
\end{equation}
Here $B$ is the full operator including all subsystems and the equation is solved for the unknown energy densities $\rho_{\Gamma}=(\rho_{\Gamma_{i}})_{i=1,..,N_{\Omega}}$, where $\rho_{\Gamma_{i}}$ denotes the (Chebyshev) coefficients of the density on $\Gamma_{i}$, the boundary of $\Omega_{i}$.

\section{\label{sec:numericimp}NUMERICAL IMPLEMENTATION}

\subsection{From SEA to DEA}
Up to now, the various representations given are all equivalent and correspond to a description of the wave dynamics in terms of the ray tracing ansatz (\ref{fpo2}). Traditional ray tracing based on sampling ray solutions over the available phase space is rather inefficient. Convergence tends to be fairly slow, especially if the absorption is low and an exponentially increasing number of long paths including multiple reflections need to be taken into account.

An SEA treatment emerges when approximating the individual operators $\mathcal{B}_{ij}$ in terms of constant functions only \cite{GT09}; using, for example, a Fourier basis this would correspond to an approximation in terms of the lowest order basis functions only. In the case of a Chebyshev basis, the integrand in (\ref{cheb-ij}) is not constant due to the presence of the weight function $W_{m}$ and an SEA treatment is obtained only after restricting the basis to $\tilde{T}^i_0$ and omitting the weights $W_{0}$. Note that when using Gauss-Chebyshev quadrature an explicit division by $W_{0}$ is required since the weight function is automatically included in the quadrature rule. In the SEA case the matrix $B_{ij}$ is one-dimensional and gives the mean transmission rate from subsystem $j$ to subsystem $i$. It is thus equivalent to the coupling loss factor used in standard SEA equations \cite{RL95}. The resulting full $N_{\Omega}$-dimensional $B$ matrix yields a set of SEA equations using the relation (\ref{lin-syst}) after mapping the boundary densities back into the interior.

An SEA approximation is justified if the ray dynamics within each subsystem are sufficiently chaotic such that a trajectory entering subsystem $j$ forgets everything about its past before exiting $\Omega_{j}$; SEA can thus be viewed as a Markov approximation of the deterministic dynamics. Thus correlations within the dynamics must decay rapidly on the time scale it takes for a typical ray to leave $\Omega_{j}$. This condition will often be fulfilled if the subsystem boundaries are sufficiently irregular, the subsystems are dynamically well separated and absorption and dissipation is small, conditions typically cited in an SEA context. In this case SEA is an extremely efficient method compared to standard ray tracing. However for subsystems with regular features, such as rectangular cavities or corridor-like elements, incoming rays are directly channeled into outgoing rays, thus rendering the Markov approximation invalid and introducing memory effects. Likewise, strong damping may lead to a significant decay of the signal before reaching the exit channel introducing geometric (system dependent) effects.

The features that cause an SEA approximation to fail as described above are incorporated into the model here by including higher order basis functions and weight functions $W_{m}$ for each subsystem boundary operator $B_{ij}$. It therefore becomes possible to resolve the fine structure of the dynamics and their correlation along with effects due to non-uniform damping over typical scales of the subsystem. The maximal number of basis functions needed to reach convergence is expected to be relatively small thus making the new method more efficient than a full ray tracing treatment, particularly when damping is low.

Representing the ray dynamics in terms of finite dimensional transition matrices corresponds to a refinement of an SEA technique, taking advantage of the efficiency of SEA but including information about the ray dynamics when necessary. It overcomes some of the limitations of SEA and puts the underlying SEA assumptions on sound foundations.

\subsection{Chebyshev DEA}\label{sec:ChebDEA}

\begin{figure}\centering
       \includegraphics[width=8cm]{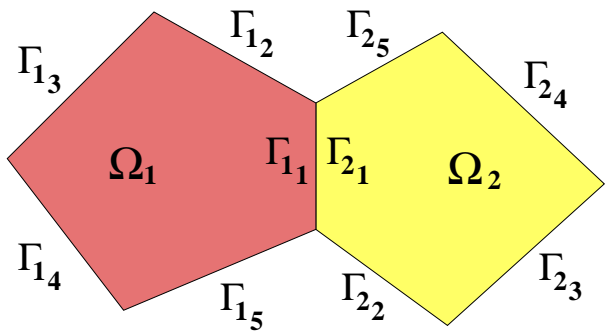}
        \caption{(Color online) A polygonal configuration with $\Gamma_{j}=\partial\Omega_{j}$ for $j=1,2$ showing further subdivision $\Gamma_{j_i}$ of the boundary. The interface is formed by the boundary sections $\Gamma_{1_1} = \Gamma_{2_1}$. A separate spatial basis function approximation is applied for each subdivision.}\label{geomref}
\end{figure}

A favourable property of the Chebyshev basis approximation compared with a Fourier basis is the flexibility it allows in the choice of approximation regions due to not requiring periodic boundary conditions for convergence. In many cases it will be advantageous to subdivide the subsystem boundaries and apply a Chebyshev basis for each subdivision. An example where this subdivision takes place at the vertices of each subsystem is shown in Fig. \ref{geomref}. A separate spatial basis expansion is then applied in a number of sections $\Gamma_{j_{i}}$, $i=1,...,N_{j}$ of the boundary $\Gamma_{j}$ of a particular subsystem $\Omega_j$,
$j=1,..,N_{\Omega}$. This leads to separate entries in the matrix representation $B$ for each boundary section $\Gamma_{j_{i}}$ and hence a larger but sparser matrix.

In order to understand why this geometric subdivision may be beneficial, we first need to consider some properties of the basis function expansions. The basis expansion of the boundary operator $\mathcal{B}$ results in density functions $\rho_{\Gamma_{i}}^{(0)}$ and $\rho_{\Gamma_{i}}$ in equation (\ref{Neum-ser}) being expanded in terms of the chosen basis functions. The convergence rate of the basis expansion therefore depends on the properties of these density functions and in particular their smoothness \cite{JB00}. The basis function approximations are carried out with respect to both position $s$ and momentum along the boundary $p_{s}$ at the endpoint of the ray, recall equation (\ref{cheb2}). A discontinuity in the normal to $\Gamma_{i}$ for some $i=1,...,N_{\Omega}$ will, in general, result in a non-differentiable initial density. It is therefore recommended to employ separate Chebyshev basis approximations with respect to $s$ for regions of $\Gamma$ split by an edge or vertex, see Fig. \ref{geomref}. It is also recommended to employ a separate approximation with respect to $s$ where there is a sudden change in boundary conditions resulting in a non-smooth initial density, for example, along an interface between two subsystems. The geometric subdivision is therefore to enable smoothness of the density functions, $\rho_{\Gamma_{i}}^{(0)}$ and $\rho_{\Gamma_{i}}$, over the regions of approximation and therefore (geometric) exponential order convergence in the Chebyshev basis expansions.

The method described above has been implemented numerically for a variety of problems with $\Omega\subset\mathbb{R}^{2}$ and homogeneous Dirichlet boundary conditions for the total wave, (that is, including the initial contribution from the source) on the outer boundary $\Gamma$. For simplicity we consider $c_i =\varrho^{-1/2}_{i}$ for all $i=1,..,N_{\Omega}$, that is, we set the adiabatic compressibility to unity. Results are compared with the numerically exact solutions obtained from state-of-the-art adaptive FEM software. A brief account of the FE method is given in the next section.

Let $s\in[0,L_{j_{i}})$ and $p_{s}\in(-k_{i},k_{i})$ parameterize the associated phase space
where $L_{j_{i}}$ is the length of $\Gamma_{j_{i}}$. In these local sub-system boundary coordinates the Chebyshev basis is given by
\begin{equation}\label{Cheb-basis}
\tilde{T}^{j_{i}}_{n}(s,p_{s})=\sqrt{\frac{2}{k_{i}L_{j_{i}}}}T_{n_{1}}(\tilde{s})T_{n_{2}}(\tilde{p}_{s}),
\end{equation}
where $\tilde{s}=(2s/L_{j_{i}})-1$ and $\tilde{p}_{s}=p_{s}/k_{i}$. The weight function $W_{n}$ is given by
\begin{equation}\label{Cheb-weight}
W_{n}(s,p_{s})=\frac{4\gamma_{n_{1}}\gamma_{n_{2}}}{\pi^{2}}\frac{1}{\sqrt{1-\tilde{s}^{2}}}\frac{1}{\sqrt{1-\tilde{p}_{s}^{2}}}.
\end{equation}
The transmission probability at the intersection of two subsystems is given by \cite{JD92}
\begin{equation}\label{rt-coeff}
w_{t}(k_{out},k_{in},\theta_{in})=\frac{4(k_{out}/k_{in})\cos(\theta_{in})\cos(\theta_{out})}{((k_{out}/k_{in})\cos(\theta_{in})+\cos(\theta_{out}))^{2}},
\end{equation}
with $\theta_{in}$ ($\theta_{out}$) the angle between the incoming (outgoing) ray and the inward normal to the boundary and $k_{in}$ ($k_{out}$) the wavenumber in the subsystem through which the incoming (outgoing) ray is traveling. Incoming and outgoing rays are related through Snell's law ($k_{in}\sin(\theta_{in})=k_{out}\sin(\theta_{out})$), and hence $\theta_{out}$ may be calculated from the other three quantities.

Given the end point $s\in\Gamma_{j_{i}}$ and the incoming momentum $p_{s}$, one can obtain the corresponding ray. With sufficient geometric knowledge of $\Gamma_{i}$ and the boundaries of its adjoining subsystems, the start point of the ray $s'$ can be determined as its intersection with one of these boundaries, say in the subset $\Gamma_{\beta_{\alpha}}$ for some $\beta_{\alpha}=1,...,N_{\alpha}$ and $\alpha=1,...,N_{\Omega}$. Once this is known it is straightforward to obtain the initial momentum $p_{s}'$. Writing out the Jacobian (\ref{chebJ}) one obtains
\begin{equation}\label{cheb3}
\begin{array}{ll} \displaystyle {B_{j_{i}\beta_{\alpha}}^{mn}=\frac{k_{\alpha}}{k_{i}}\int_{k_{j_{i}}^{min}}^{k_{j_{i}}^{max}}\int_{0}^{L_{j_{i}}}} & {\displaystyle\hspace{-2mm} W_{m}(s,p_{s})\tilde{T}_{m}^{j_{i}}(s,p_{s})\tilde{T}_{n}^{\beta_{\alpha}}(s',p_{s}')\times}\hspace{2mm}\\{} & {\displaystyle w_{j_{i}\beta_{\alpha}}(s',p_{s}')ds\:dp_{s},}
\end{array}
\end{equation}
where $w_{j_{i}\beta_{\alpha}}(s',p_{s}')=\exp(-\mu_{i} L)w^{\Gamma}_{j_{i}\beta_{\alpha}}(s',p_{s}')$. Here $\mu_{i}$ is the damping coefficient in $\Omega_{i}$ as before, $L$ is the length of the trajectory from $s'$ to $s$ and the reflection/ transmission coefficients are
$$w^{\Gamma}_{j_{i}\beta_{\alpha}}(s',p_{s}')=\left\{\begin{array}{ll}
{\delta_{i\alpha}} & {\mathrm{if}\:s\notin\Gamma_{i\alpha}}\\ {\delta_{i\alpha}+(-1)^{\delta_{i\alpha}}w_{t}(k_{i},k_{\alpha},\theta_{i})} & {\mathrm{if}\:s\in\Gamma_{i\alpha}.}\end{array}\right.$$
Here, $\Gamma_{i\alpha}$ denotes any subset of $\Gamma_{\alpha}$ forming the intersection of two subsystems (that is, for $i\neq\alpha$, $\Gamma_{i\alpha}=\Gamma_{i}\cap\Gamma_{\alpha}$) and $\delta_{i\alpha}$ is the Kronecker delta. Also $k_{j_{i}}^{min}=k_{i}\sin(\theta_{j_{i}}^{min})$ and $k_{j_{i}}^{max}= k_{i}\sin(\theta_{j_{i}}^{max})$ are the minimum and maximum values of $p_{s}$, respectively, where $\theta_{j_{i}}^{min}$, $\theta_{j_{i}}^{max}\in(-\pi/2,\pi/2)$ are the angles between the inward normal to $\Gamma_{j_{i}}$ and the rays from $s'$ to each of the ends of $\Gamma_{j_{i}}$.

All DEA/ SEA computations are performed on a desktop PC with a dual core 2.83 GHz processor using C++ with Diffpack ({\tt www.diffpack.com}).

\subsection{\label{sec:FEM}$hp$-adaptive discontinuous Galerkin finite element method}
In the Sec.\ ref{sec:computations}, we will compare DEA results twith numerically exact solutions of the wave equation (\ref{wave-eq}). Discontinuous Galerkin (DG)  methods have become increasingly popular for elliptic problems in recent years  \cite{brezzi} and is also our method of choice for
calculating the Green function.The main reason for this interest in DG methods is that
allowing for discontinuities across elements gives extraordinary flexibility
in terms of mesh design and choice of shape functions.
Additionally, $hp$-adaptive DG methods, which are based on locally refined
meshes and variable approximation orders, achieve tremendous gains in computational efficiency for challenging problems \cite{HartmannHouston,HartmannHouston-ns, houstonschotzauwihler, houston1}.

In the following, we consider the wave equation
\begin{equation}\label{wave-eq1}
(c_i^2 \Delta +\tilde{\omega}_{i}^{2})\tilde{G}(r,r_{0};\omega)=-\delta(r-r_{0}), \hspace{5mm}i=1,...,N_{\Omega},
\end{equation}
with $\tilde{\omega}_i = \omega+i\mu_{i}c_{i}/2$ and $r\in\Omega_i \subset\mathbb{R}^2$. The Green function $\tilde{G}$
is related to $G$ in equation (\ref{wave-eq}) by $G = c_0^2 \tilde{G}$, where $c_0$ is the wave velocity in the subsystem $\Omega_0$ containing the source point $r_{0}$.

To drive the $hp$-adaptivity we use an explicit energy norm \textit{a posteriori} error estimator inspired by Refs. \cite{dominik, giani_dominik}. We apply a simple fixed-fraction strategy on the error estimator to mark the elements to adapt. For each marked element, the choice of whether to locally refine it or vary its approximation order is made by estimating the decay of the coefficients in an $L^2$-orthogonal polynomial expansion in order to test the local analyticity of the solution in the interior of the element \cite{melenk}. The details of the implementation of the method can be found in Appendix~\ref{app:dg}.

All finite element simulations have been carried out using AptoFEM ({\tt www.aptofem.com})
on a parallel machine.

\section{\label{sec:computations}COMPUTATIONAL RESULTS}
\subsection{Coupled two-cavity systems}
A variety of two-cavity systems are considered as in Ref. \cite{GT09} and are shown in Fig. \ref{twodomplots}. The coordinates of the vertices and source points of these systems are given in Appendix \ref{coordapp}.
We compare our DEA results with standard SEA calculations; relevant parameters such as the area of each subsystem have been listed in the Appendix \ref{coordapp}. The SEA coupling loss factors are obtained from the transmission coefficients (\ref{rt-coeff}) integrated over the
wavenumber $k_{in}$ of the incoming rays. The attenuation is given by the damping factor $\mu$; in the following we will assume
hysteretic damping with $\mu_{i}=\omega\eta / (2c_{i})$ for $i=1,2$.

Configuration A features irregular shaped well separated pentagonal subsystems and thus SEA is expected to work well. In configuration B the size of the interface between the subsystems is increased reducing their dynamical separation and therefore the applicability of SEA. Configuration C includes a rectangular left-hand subsystem channeling rays out of the subsystem and introducing long-range correlations in the dynamics. In addition, the source is further from the intersection of the two subsystems. SEA is thus not expected to work well for this configuration. Note that SEA results are in general insensitive to the position of the source, whereas actual trajectory calculations may well depend on the exact position.

\begin{figure}\centering
       \includegraphics[width=9cm]{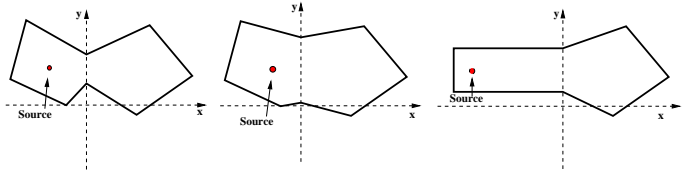}
        \caption{Coupled two-domain systems: configurations A, B and C, respectively.}\label{twodomplots}
\end{figure}

Finite basis sets have been employed with $n_{1},n_{2}=0,..,N$, which gives rise to
matrices of size $\mathrm{dim}B=(N+1)^{2}(N_{1}+N_{2})$,
with basis functions of the same order for position and momentum in both subsystems.
Note that $N_i, i=1,2$ denotes the number of subdivisions of the boundary of subsystem $i$ as defined in Section \ref{sec:ChebDEA}.
Energy distributions have been studied as a function of the frequency with a hysteretic damping factor $\eta=0.01$. Here and in the remainder of this work the subsystems are numbered $1,...,N_{\Omega}$ from left to right.

\begin{figure}\centering
       \includegraphics[width=9cm]{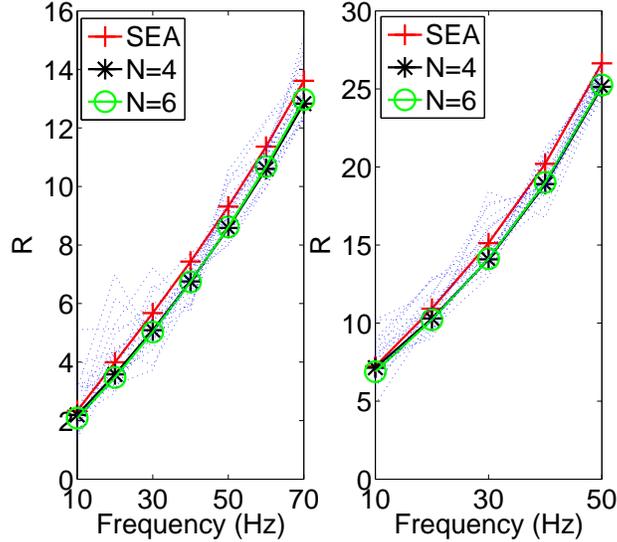}
         \caption{(Color online) Ratio of total energies $\textsf{R}=\|\tilde{G}_{1}\|^{2}/\|\tilde{G}_{2}\|^{2}$ in configuration A with $c_{1}=c_{2}=1$ (left) and $c_{1}=0.5$, $c_{2}=1$ (right). Dotted lines show FEM results computed within $\pm 5$Hz of the DEA results.}\label{ratplot1}
\end{figure}

Fig.\ \ref{ratplot1} shows approximations of the ratios of the total energy in each subsystem $\|\tilde{G}_{1}\|^{2}/\|\tilde{G}_{2}\|^{2}$ in configuration A where
\begin{equation}\label{ennorm}
\|\tilde{G}_{i}\|^{2}:=\int_{\Omega_{i}}|\tilde{G}(r,r_{0};\omega)|^{2}dr,\hspace{2mm}i=1,...,N_{\Omega}.
\end{equation}
The left subplot shows the results with $c_1=c_2=1 \mathrm{ms}^{-1}$ and the right with $c_1=0.5\mathrm{ms}^{-1}$ and $c_2=1\mathrm{ms}^{-1}$. The dotted lines represent solutions computed using the FEM (described in section \ref{sec:FEM}) at an equi-spaced range of frequencies within $\pm 5$Hz of the frequencies used for the SEA and DEA computations. Note that the damping is fixed to the value employed for the central (SEA/DEA) frequency. In the right-hand plot the results only go up to $50$Hz due to the high computational cost of the FEM code for the large wavenumbers in subsystem 1. Fast convergence with increasing basis size is evident in each case. It is clear that SEA works reasonably well for configuration A since the SEA prediction is close to that from DEA and within the range of FEM solution values. In particular the SEA prediction is good for low damping values (that is, low frequencies). At closer inspection one notices also that the divergence from the DEA result is smaller in the case when $c_1\neq c_2$, thus demonstrating an increase in the dynamical separation between subsystems in this case. Comparing SEA with the $N=6$ case between $10$Hz and $50$Hz, the SEA results differ from DEA by between $8\%$ and $15\%$ when $c_1=c_2=1 \mathrm{ms}^{-1}$, but only by between $5\%$ and $7\%$ when $c_1=0.5 \mathrm{ms}^{-1}$ and $c_2=1 \mathrm{ms}^{-1}$.

\begin{figure}\centering
       \includegraphics[width=9cm]{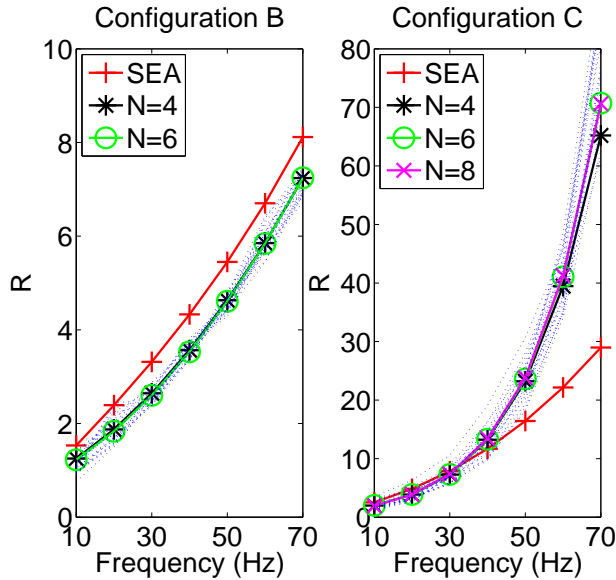}
        \caption{(Color online) Ratio of total energies $\textsf{R}=\|\tilde{G}_{1}\|^{2}/\|\tilde{G}_{2}\|^{2}$ in configuration B (left) and configuration C (right). Dotted lines show FEM results computed within $\pm 5$Hz of the DEA results.}\label{ratplot2}
\end{figure}

Figure \ref{ratplot2} shows approximations of the ratios of the total energy in each subsystem in configurations B and C with $c_1=c_2=1 \mathrm{ms}^{-1}$. The dotted lines are as before and fast convergence with increasing basis size is again evident, although a slightly higher order was required for configuration C to capture the exponential decay due to dissipation along the rectangular cavity. As expected, the SEA prediction diverges from both the FEM and DEA predictions in configurations B and C. Comparing SEA with the $N=6$ case for configuration B between $10$Hz and $50$Hz (for consistency with the data for configuration A), the SEA results differ from DEA by between $18\%$ and $29\%$. For configuration C the deviation is about $30\%$ in the range 10Hz to 50 Hz, although this grows considerably larger if one includes the data for $60$ and $70$Hz.

\subsection{Complex built-up systems}
\begin{figure}\centering
       \includegraphics[width=9cm]{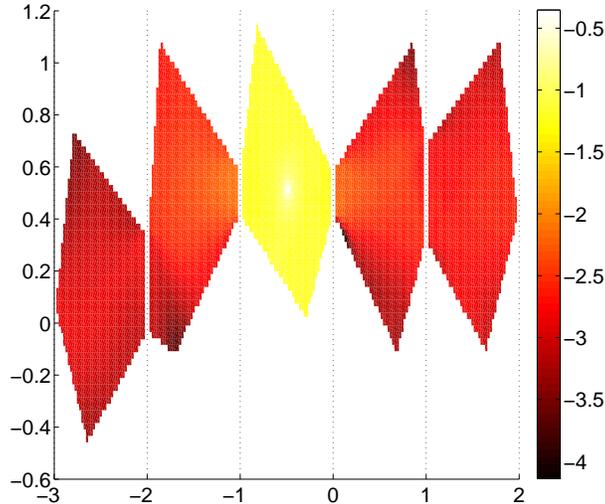}
        \caption{(Color online) Distribution of $\log_{10}(|\tilde{G}|^{2})$ in a five cavity system.}\label{fiveplot1}
\end{figure}
In this section the versatility and efficiency of the Chebyshev approximation with Gauss-Chebyshev quadrature are demonstrated by considering a complex built-up system. A configuration of five coupled acoustic cavities is considered as shown in Figure \ref{fiveplot1}. The coordinates of the vertices and source point are given in Appendix \ref{coordapp}. The solution in the interior of each cavity is plotted and the source point in the central cavity is clearly evident. The subsystems are each convex polygonal regions as before, the jagged appearance of the boundary is a result of the plotted region being formed of the largest regular grid fitting strictly inside the boundary of each subsystem. The wave velocities are taken to be $c_1=c_2=c_4=c_5=1 \mathrm{ms}^{-1}$ and $c_3=0.5 \mathrm{ms}^{-1}$. Figure \ref{fiveplot1} shows the DEA approximation of the distribution of $\log_{10}(|\tilde{G}|^{2})$ throughout the system with $N=8$. The logarithm of the solution is taken since a large range of values are present between the peak of the source and the extremal subsystems. The plot is for the $10$Hz case with the same frequency and damping correspondence as in the earlier two-cavity configurations. DEA clearly gives much more detailed spatial information about the wave energy distributions than SEA, which assumes a constant density in each subsystem. In particular here one can see local variations close to subsystem interfaces and a drop in the intensity as one moves away from the source. Note that more energy flows into the far right subsystem as compared to the far left subsystem due to there being a direct channel for the energy to travel along to the right of the source, but not to the left.
\begin{figure*}\centering
     \includegraphics[width=16cm]{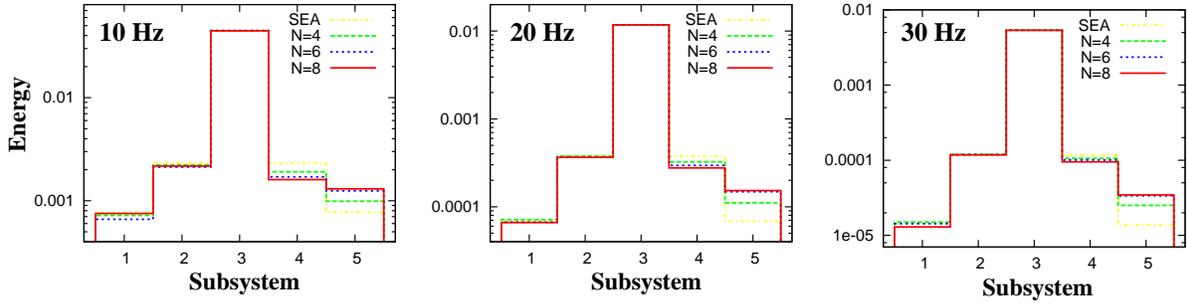}
             \caption{(Color online) A plot of $\|\tilde{G}_{i}\|^{2}$ for subsystems $\Omega_i$, $i=1,..,5$ at three different frequencies.}\label{fiveplot2}
\end{figure*}

Figure \ref{fiveplot2} shows approximations of $\|\tilde{G}_{i}\|^{2}$ for $i=1,..,5$ computed using both SEA and DEA up to $N=8$. The three figures represent the total energy (actually $c_{0}^{4}=1/16$ multiplied by this quantity since $\tilde{G}$ rather than $G$ has been computed) obtained for each subregion at three different frequencies and thus damping levels; using the same parameters as before the 10, 20 and 30 Hz cases are shown from left to right. Note that results are shown on a logarithmic scale and that the overall amplitude decreases with increasing frequency due to increased damping. Here a comparison with FEM simulations is not considered due to the high computational cost associated with computing high and multi-frequency solutions over large domains. We can deduce that SEA is working very well in subsystems $1$ to $3$ due to the level of agreement with the DEA calculations. Very few trajectories can move from the source directly into subsystem 1 and multiple scattering events lead to a local equidistribution, that is, incoming and outgoing rays in subsystem 2 are uncorrelated. The situation is different for subsystems 4 and 5 where the influence of the direct channel from the source to subsystem 5 becomes important. Compared to the SEA result, the DEA calculations show a noticeable increase in the values of $\|\tilde{G}_{5}\|^{2}$ and a slight decrease in the values of $\|\tilde{G}_{4}\|^{2}$, that is, more energy reaches subsystem 5 than predicted by a Markov approximation of the dynamics. The DEA calculations again appear to converge reasonably quickly between $N=6$ and $N=8$. It is also evident that SEA works best for the lower damping values here, which is consistent with our observations for the two-cavity configurations.

\section{CONCLUSIONS}
Dynamical energy analysis has been reformulated in terms of a Chebyshev basis expansion allowing a greater degree of flexibility in the design of the method and more efficient coding. A comparison of the numerical results with finite element method computations showed that dynamical energy analysis is robust and retains accuracy in cases when SEA fails. Examples where the wave velocity varies between subsystems were also considered for the first time using DEA. An extension to built-up systems, facilitated by the efficient Chebyshev basis reformulation of DEA, showed that DEA is a flexible, robust and efficient method for estimating energy density distributions at high frequencies.

\section*{Acknowledgments}
Support from the EPSRC (grant EP/F069391/1) and the EU (FP7IAPP grant MIDEA) is gratefully acknowledged.

The authors also wish to thank Inutech Gmbh, N\"{u}rnberg for Diffpack guidance and licences, and Dmitrii Maksimov for carefully reading the manuscript.

\appendix
\section{Implementation of the DG method}\label{app:dg}

In order to compute an approximation of $\tilde{G}$ in \eqref{wave-eq1},
we partition the domain $\Omega$ with shape-regular meshes ${\mathcal T}_h$ formed by  open triangles
$\{K\}_{K\in{\mathcal T}_h}$. We assume that in the interior of each element $K\in{\mathcal T}_h$, the wave velocity and the damping are constant, such values are denoted in the interior of $K$ by $c_K$ and $\mu_K$.
Further, each element $K$ can
then be affinely mapped onto the generic reference element
$\widehat{K}$. The diameter of an element $K\in{\mathcal T}_h$ is denoted by
$h_K$. Due to our assumptions on the meshes, these diameters are of
bounded variation, that is, there is a constant $b_1\ge 1$ such
that
\begin{equation}
\label{eq:boundedvariationh}
b_1^{-1} \leq h_{K}/h_{K'} \leq b_1,
\end{equation}
whenever $K$ and $K^\prime$ share a common edge. We store the
elemental diameters in the mesh size vector $\uu{h}=\{h_K\,:\,K\in{\mathcal T}_h\}$. Similarly, we associate with
each element $K\in{\mathcal T}_h$ a polynomial degree $p_K\geq 1$ and
define the degree vector $\uu{p}=\{p_K\,:\,K\in{\mathcal T}_h\}$. We
assume that $\uu{p}$ is of bounded variation as well, that is, there
is a constant $b_2 \ge 1$ such that
\begin{equation}
\label{eq:boundedvariationp}
b_2^{-1} \leq p_{K}/p_{K'} \leq b_2,
\end{equation}
whenever $K$ and $K^\prime$ share a common edge.

For a partition ${\mathcal T}_h$ of $\Omega$
and a degree vector~$\uu{p}$,
we define the $hp$-version discontinuous Galerkin finite
element space $V_h$ of complex valued functions by
\begin{equation}
\label{eq:hpspace}
V_h=\{\, v\in L^2(\Omega)\, :\, v|_K \in {\mathcal P}_{p_K}(K),\
K\in {\cal T}_h\,\}.
\end{equation}
Here, ${\mathcal P}_{p_K}(K)$ is the space
of polynomials on $K$ of total degree less than or equal to
$p_K$.

Next, we define some trace operators that are required for the DG methods. To this end, we denote by ${\mathcal
E}_\calI({\mathcal T_h})$ the set of all interior edges
and by ${\mathcal E}_\Gamma({\mathcal T}_h)$ the set of all boundary edges of
the partition ${\mathcal
T}_h$. Furthermore, we define ${\mathcal E}({\mathcal T}_h)={\mathcal
E}_\calI({\mathcal T}_h) \cup {\mathcal E}_\Gamma({\mathcal T}_h)$. The
boundary $\partial K$ of an element $K$ and the sets $\partial K
\setminus \Gamma$ and $\partial K \cap \Gamma$ will be identified in a
natural way with the corresponding subsets of ${\mathcal E}({\mathcal
T}_h)$.

Let $K^+$ and $K^-$ be two adjacent elements of~ ${\mathcal T}_h$, and $\kappa\in{\mathcal
E}_{\mathcal I}({\mathcal T}_h)$ given by $\kappa=\partial
K^+\cap\partial K^-$. Furthermore, let $v$ be a complex scalar-valued function, that is smooth inside
each element~$K^\pm$. By $v^\pm$, we denote the traces of
$v$ on $\kappa$ taken from within the interior of $K^\pm$,
respectively. Then, the weighted average of the diffusive flux $c^{2}\nabla_hv$ along $
\kappa\in{\mathcal
E}_{\mathcal I}({\mathcal T}_h)$ is given by
$$
\mvl{c^{2}\nabla_hv}=\frac{c_{K^-}^{2}c_{K^+}^{2}\nabla_hv^+
+c_{K^+}^{2}c_{K^-}^{2}\nabla_hv^-}
{c_{K^+}^{2}+c_{K^-}^{2}}.
$$ Similarly, the jump of $v$ across
$\kappa\in{\mathcal
E}_{\mathcal I}({\mathcal T}_h)$ is given by
$$
\jmp{v} =v^+\,\uu{n}_{K^+}+v^-\,\uu{n}_{K^-},
$$
where we denote by $\uu{n}_{K^\pm}$ the unit outward
normal vector of $\partial K^\pm$, respectively.

On a boundary edge $\kappa\in{\mathcal E}_\Gamma({\mathcal
T}_h)$, we set $\mvl{c^{2}\nabla_hv}=c^{2}\nabla_hv$ and
$\jmp{v}=v\uu{n}$ ,
with~$\uu{n}$ denoting the unit outward normal
vector on the boundary $\Gamma$.

For a mesh ${\mathcal T}_h$ on $\Omega$ and a polynomial degree vector
$\uu{p}$, let $V_h$ be the $hp$-version finite element space defined
in \eqref{eq:hpspace}. We consider the (symmetric) weighted interior penalty
discretization \cite{ern} of~\eqref{wave-eq1}:
find $\tilde{G}_{h}\in V_h$ such that
\begin{equation}
\label{eq:DGFEM}
A_h(\tilde{G}_{h},v)\ =\ F_h(v)\ ,\quad \mbox{for all } v\in V_h\ ,
\end{equation}
where
\begin{align*}
A_{h}(u,v)
&:=\sum_{K\in{\mathcal T}_h}\int_K\, c_K^{2}\nabla_h u\cdot\nabla_h
\overline{v} -(\omega+i \mu_K c_K /2)^2u \overline{v}\,d\uu{r}\\
& \quad-\sum_{\kappa\in{\mathcal E}({\mathcal T}_h)}\int_\kappa\,
\big(\overline{\mvl{c^{2}\nabla_h v}}\cdot\jmp{u}+\mvl{c^{2}\nabla_h u}\cdot \overline{\jmp{v}}\big) \, ds\\
& \quad+\sum_{\kappa\in{\mathcal E}({\mathcal T}_h)}\int_{\kappa}\,\cc \,
\jmp{u}\cdot \overline{\jmp{v}}\, ds,\\
F_h(v) &:= \int_\Omega\, \delta(r-r_0) \overline{v}\,d\uu{r}\ ,
\end{align*}
and $\nabla_h$ denotes the element wise gradient operator.
Since DG methods allow for discontinuities in the finite
element approximation, the supports of the shape functions never extend
to more than one element. A consequence is that the stencil is minimal in the sense that
each element communicates only with its direct neighbors and the
communication happens across the edges of the mesh where the functions in
the finite element space are not continuous. This makes it natural to include the last two terms in the
definition of $A_h(\cdot,\cdot)$, which control the average and the flux of discontinuous functions across the
edges. In particular the stability of the method is guaranteed by the
third term, which penalizes the discontinuities across the edges. So the second and the third terms
are peculiar to the DG method of choice, in contrast to the first term which depends on the PDE under consideration.
Furthermore, the function $\cc\in L^\infty({\mathcal E}({\mathcal
T}_h))$ is the discontinuity stabilization function that is chosen as
follows: we define the functions $\hh\in L^\infty({\mathcal
E}({\mathcal T}_h))$ and $\pp\in L^\infty({\mathcal E}({\mathcal
T}_h))$ by
\begin{equation*}
\begin{split}
\hh(\uu{r}) &:=
\begin{cases}
\min(h_K,h_{K'}), & \uu{r}\in \kappa\in{\mathcal
E}_{\calI}({\mathcal T}_h),\
\kappa=\partial K\cap\partial K',\\
h_K, & \uu{r}\in \kappa\in{\mathcal E}_\Gamma({\mathcal T}_h),
\ \kappa\in\partial K\cap\Gamma,
\end{cases}\\
\pp(\uu{r}) &:=
\begin{cases}
\max(p_K,p_{K'}), & \uu{r}\in
\kappa\in{\mathcal E}_{\calI}({\mathcal T}_h),
\ \kappa=\partial K\cap\partial K',\\
p_K, &  \uu{r}\in \kappa\in{\mathcal E}_\Gamma({\mathcal T}_h),\
\kappa\in \partial K\cap\Gamma,
\end{cases}
\end{split}
\end{equation*}
and set
\begin{equation}
\label{eq:IPdef}
\cc=\gamma \frac{c_K^{2}c_{K'}^{2}}{c_K^{2}+c_{K'}^{2}}\frac{\pp^2}
{\hh},
\end{equation}
with a parameter $\gamma>0$ that is independent of $\uu{h}$,
$\uu{p}$, $c_K$ and $c_{K'}$. The definition of $\cc$ is equivalent to an $hp$-version
of the weighted penalty parameter in Ref. \cite{ern}.
The definitions of $\hh(\uu{r})$, $\pp(\uu{r})$ and $\cc$ are designed to work with $hp$-adaptivity where adjacent elements of very different sizes and orders are possible. In order to keep the method stable under the adaptation process it is necessary to
penalize more discontinuities across the smaller edges and also across
higher order elements. This can be seen straightaway from
\eqref{eq:IPdef}.

\section{Geometric and SEA data}\label{coordapp}
\begin{table}\caption{Coordinates for the numerical examples in Section V}\label{coord_tab} 
 \begin{tabular}{cccc}
  {Configuration} & {Vertex} & {$x$} & {$y$}\\
  \hline
  {A} & {1} & {1.4564} & {0.40381} \\
  {}  & {2} & {0.87} &   {1.1027}\\
  {}  & {3} & {0.0} &   {0.6993}\\
  {}  & {4} & {-0.83} & {1.1720}\\
  {}  & {5} & {-1.048} & {0.3582}\\
  {}  & {6} & {-0.28} & {0.0}\\
  {}  & {7} & {0.0} & {0.2993}\\
  {}  & {8} & {0.69} & {-0.1328}\\
  \hline
  {B} & {1} & {1.4564} & {0.40381} \\
  {}  & {2} & {0.87} &   {1.1027}\\
  {}  & {3} & {0.0} &   {0.9493}\\
  {}  & {4} & {-0.83} & {1.1720}\\
  {}  & {5} & {-1.048} & {0.3582}\\
  {}  & {6} & {-0.28} & {0.0}\\
  {}  & {7} & {0.0} & {0.0493}\\
  {}  & {8} & {0.69} & {-0.1328}\\
  \hline
  {C} & {1} & {1.4564} & {0.40381} \\
  {}  & {2} & {0.87} &   {1.1027}\\
  {}  & {3} & {0.0} &   {0.7993}\\
  {}  & {4} & {-1.503} & {0.7993}\\
  {}  & {5} & {-1.503} & {0.1993}\\
  {}  & {6} & {0.0} & {0.1993}\\
  {}  & {7} & {0.69} & {-0.1328}\\
  \hline
  {5 Cavity} & {1} & {2.0} & {0.41} \\
  {} &  {2} & {1.8} & {1.1027} \\
  {}  & {3} & {1.0} & {0.6993}\\
  {}  & {4} & {0.87} & {1.1027}\\
  {}  & {5} & {0.0} & {0.5993}\\
  {}  & {6} & {-0.83} & {1.1720}\\
  {}  & {7} & {-1.0} & {0.5993}\\
  {}  & {8} & {-1.87} & {1.1027}\\
  {}  & {9} & {-2.0} & {0.3493}\\
  {}  & {10} & {-2.8} & {0.7527}\\
  {}  & {11} & {-3.0} & {0.06}\\
  {}  & {12} & {-2.65} & {-0.4828}\\
  {}  & {13} & {-2.0} & {-0.0507}\\
  {}  & {14} & {-1.69} & {-0.1328}\\
  {}  & {15} & {-1.0} & {0.3993}\\
  {}  & {16} & {-0.28} & {0.0}\\
  {}  & {17} & {0.0} & {0.3993}\\
  {}  & {18} & {0.69} & {-0.1328}\\
  {}  & {19} & {1.0} & {0.2993}\\
  {}  & {20} & {1.65} & {-0.1328}\\
 \end{tabular}
 \end{table}

 \begin{table}\caption{Subsystem areas and interface lengths for the numerical examples in Section V} \label{SEAdet}
 \begin{tabular}{cccc}
  {Configuration} & {Subsystem} & {Area (to 4sf)} & {Interface length (with next subsystem)}\\
  \hline
  {A} & {1} & {0.7639} & {0.4} \\
  {}  & {2} & {1.064} &   {-}\\
   \hline
  {B} & {1} & {0.9027} & {0.9} \\
  {}  & {2} & {1.259} &   {-}\\
  \hline
  {C} & {1} & {0.9018} & {0.6} \\
  {}  & {2} & {1.142} &   {-}\\
   \hline
  {5 Cavity} & {1} & {0.7694} & {0.4} \\
  {} &  {2} & {0.7713} & {0.2} \\
  {}  & {3} & {0.6860} & {0.2}\\
  {}  & {4} & {0.7398} & {0.4}\\
  {}  & {5} & {0.7694} & {-}\\
 \end{tabular}
 \end{table}
For completeness the coordinates of the vertices for the systems treated numerically in the paper are detailed in Table \ref{coord_tab}. For consistency with the units quoted in the paper the distances should be considered in metres. In each configuration the vertices are ordered anticlockwise starting from the vertex with the maximum $x$-coordinate. For configurations A and B the source point is $(-0.4, 0.5)$, for configuration C it is $(-1.4,0.4993)$ and for the five plate configuration the source is located at $(-0.5, 0.4993)$. In configurations A, B and C the interface between the subsystems is located on the intersection with the line $x=0$. For the five cavity configuration the interfaces are on the intersections with the lines $x=-2$, $x=-1$, $x=0$ and $x=1$. Table \ref{SEAdet} gives additional geometric parameters useful for SEA verification. Note that in a DEA framework the SEA coupling loss factors are computed using the transmission law (\ref{rt-coeff}) and details of the damping, frequency and wave velocity are provided in Section \ref{sec:numericimp}.

\end{document}